\documentclass[review,3p,times, authoryear]{elsarticle}

\usepackage{amssymb}
\usepackage{amsmath}
\usepackage{xcolor}
\usepackage[normalem]{ulem}
\usepackage{url}

\journal{Applied Food Research}

\long\def\revadd#1{{\color{black}#1}}
\long\def\revdel#1{}

\begin{document}

\begin{frontmatter}



\title{TastePrint: A 3D Food Printing System for Layer-wise Taste Distribution via Airbrushed Liquid Seasoning} 


\author{Yamato Miyatake, Parinya Punpongsanon} 

\affiliation{organization={Saitama University},
            addressline={}, 
            city={Saitama},
            postcode={338-8570}, 
            state={Saitama},
            country={Japan}}

\begin{abstract}
3D food printing enables the customization of food shapes and textures, but typically produces uniform taste profiles due to the limited diversity of printable materials.
We present TastePrint, a 3D food printing system that achieves layer-wise spatial taste distribution by dynamically applying liquid seasonings with a programmable airbrush during fabrication.
The system integrates (1) a graphical user interface (GUI) that allows users to import 3D models, slice them into layers, and specify seasoning channels, spray positions, and intensities, and (2) a customized 3D food printer equipped with a multi-nozzle spray mechanism.
\revadd{We evaluated the system through technical experiments quantifying spray resolution and deposition accuracy, a minimal sensory discrimination study on taste localization, and an exploratory formative user-feedback study involving three home cooks.}
\revadd{The spray-resolution model achieved $R^2 = 0.86$, and the spray-amount model achieved $R^2 = 0.99$. The filter-paper calibration showed broad consistency with measurements obtained on edible mashed-potato samples.
In the sensory discrimination study, participants identified the centralized seasoning pattern as more localized in 27 of 40 trials (67.5\%).}
\revadd{These findings indicate that TastePrint can provide repeatable hardware-level control over seasoning placement and quantity while offering initial evidence that spatial taste arrangement can remain perceptually meaningful after fabrication.}
\end{abstract}

\begin{graphicalabstract}
\includegraphics[width=\textwidth]{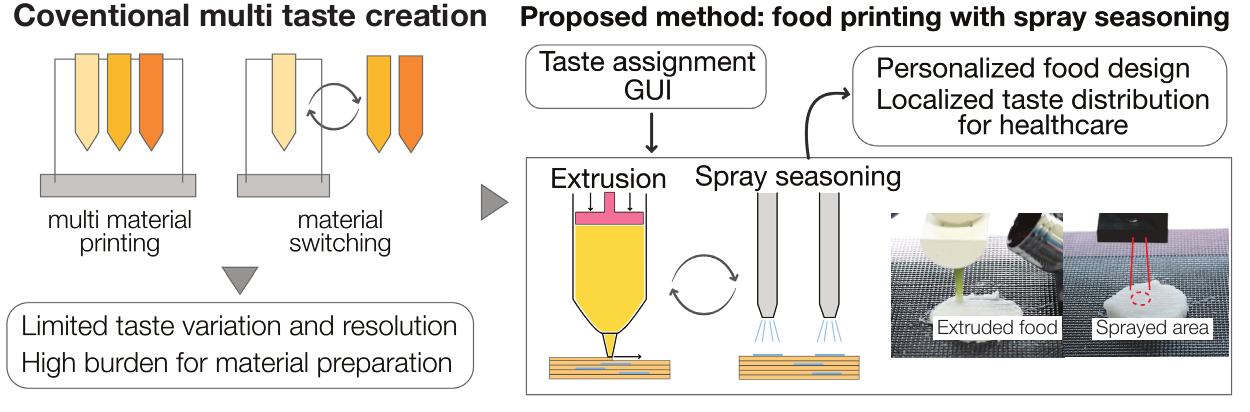}
\end{graphicalabstract}

\begin{highlights}
\item 3D food printing system achieves layer-wise taste control using airbrushing
\item GUI enables layer-specific design of internal taste distributions
\item Spray-resolution and deposition tests established technical feasibility, and a minimal sensory study provided initial evidence of perceptible spatial taste control
\end{highlights}


\end{frontmatter}



\section{Introduction}
\label{introduction}
Additive manufacturing, or 3D printing, has expanded opportunities for product design by enabling the fabrication of complex geometries and customized structures~\citep{ComPot:2022:Praveena-IG, AddRev:2023:Bhatia-PZ}. 
In the food sector, \emph{3D food printing} extends these capabilities to the culinary domain, allowing precise control of shape, size, and internal structure for personalized meals~\citep{ExtCon:2018:Sun-CJ, 3DPro:2022:Demei-RX}. 
Major techniques include extrusion~\citep{DuaGel:2018:Liu-AQ, ExtPos:2022:Hussain-GW}, which deposits food pastes through a nozzle; binder jetting~\citep{JetMat:2020:Vadodaria-PW, CrePri:2022:Zhu-GJ}, which applies liquid binders to powder beds; and inkjet printing~\citep{DevPri:2019:Suzuki-GZ, InkFoo:2023:Burkard-ZE}, which dispenses droplets of edible material.
Among these, extrusion-based printing is the most common because it is compatible with a wide range of food inks and can form intricate 3D geometries layer by layer~\citep{3DDev:2019:Voon-EY}.

While geometric customization has been extensively explored, taste control remains limited. Taste is a key determinant of food quality~\citep{FlaRev:2022:Wang-NE}, yet current printing workflows largely produce foods with uniform taste.
Although taste perception involves multisensory factors such as aroma, texture, and visual cues~\citep{ModCue:2005:Zampini-NZ, LevPer:2020:Vi-OB,EatVR:2023:Weidner-XE}, the spatial distribution and concentration of seasonings remain dominant drivers of perceived taste.
Conventional culinary practice often creates variety through staged seasoning and composition~\citep{CulArt:2022:Lee-RK, CulDis:2004:Gustafsson-LO}, whereas food scientists have sought to embed flavors into printable inks by adjusting ingredient ratios or adding taste compounds~\citep{ExtPos:2022:Hussain-GW, RhePri:2022:Cheng-EA, EdiRev:2024:Hakim-PQ, ChaPro:2025:Domzalska-DH}. However, these approaches typically yield homogeneous taste distributions, limiting the expressive and sensory diversity of 3D-printed foods.
Achieving localized, layer-dependent taste variation remains a core challenge for advancing personalized and multisensory eating experiences~\citep{EnhSuc:2010:Mosca-CO, SalChl:2010:Noort-OI, InkFoo:2023:Burkard-ZE, SenSta:2021:Fahmy-GX}.
\revdel{Beyond enhancing hedonic enjoyment, spatial control of taste also has nutritional implications. Concentrating seasonings in localized regions--—rather than dispersing them uniformly---can preserve perceived flavor intensity while reducing the total use of salt or sugar.
Such selective taste modulation offers a pathway toward healthier yet flavorful meals, highlighting the need for printing methods capable of fine-grained taste placement.}

Previous studies have explored this direction through multi-material 3D printing systems~\citep{DuaGel:2018:Liu-AQ, SenSta:2021:Fahmy-GX, MulMix:2025:Fujiwara-EX, ColSur:2025:Mendoza-Bautista-CA, ScaPro:2025:Pan-WQ}.
Multi-head printers can deposit different food inks, each containing distinct taste agents, to create localized flavor regions.
For example, \citet{SenSta:2021:Fahmy-GX} achieved spatial taste variation using a dual-head printer with sweet and salty inks, while \citet{MulMix:2025:Fujiwara-EX} demonstrated four-channel taste control with a quad-screw nozzle.
However, these approaches face two major limitations:
(1) Scalability constraints—the achievable taste complexity is limited by the number of print heads and ink reservoirs; and
(2) Operational inefficiency—preparing multiple taste-specific inks requires extensive setup, increases ingredient waste, and reduces flexibility in recipe design.
These constraints hinder broader adoption and motivate the development of a more flexible, post-extrusion seasoning method for scalable, dynamic taste modulation in 3D-printed food.

\begin{figure*}[t]
    \centering
    \includegraphics[width=\textwidth]{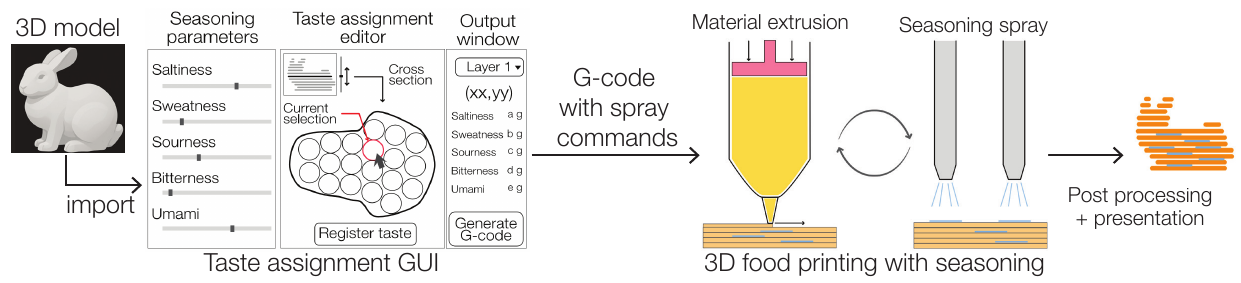}
    \caption{Overview of the TastePrint system. Users first create a customized g-code file using the GUI, where they can import a 3D model, slice it into layers, and assign taste distributions for each layer.
    The TastePrint system creates 3D-printed food with spatially controlled taste distributions by dynamically spraying liquid seasonings during the printing process.
    The printed food can be consumed directly or further processed (e.g., baking, frying).
    }
    \label{fig:system_overview}
\end{figure*}

To overcome these constraints, we propose \textbf{TastePrint}, a novel 3D food printing system that achieves spatial taste customization by dynamically airbrushing liquid seasonings during the printing process.
Instead of preparing multiple taste-specific inks, TastePrint separates geometry formation from taste modulation, enabling the creation of intricate, scalable taste patterns using a single base material.
This approach brings programmable taste modulation directly into the layer-by-layer fabrication workflow~\citep{TasMod:2023:Brooks-ZG, TTTSpr:2021:Miyashita-ES, TTTBev:2022:Miyashita-RO}.

\revdel{The TastePrint system integrates three key components: 1. a multi-nozzle airbrushing mechanism synchronized with extrusion-based printing, 2. a graphical user interface (GUI) for layer-wise taste assignment and visualization, and 3. a custom G-code generator that merges extrusion and spray commands.
Together, these elements allow users to design and fabricate foods with controllable, spatially varied taste profiles.

We validated the system through both technical and usability evaluations.
Technical experiments quantified spray resolution and deposition accuracy under varying airbrush parameters, while a user study with home cooks assessed interface usability and creative engagement.
Results showed that TastePrint achieved precise layer-wise seasoning control and was perceived as intuitive and creatively inspiring.}
\revadd{The TastePrint prototype integrates a spray-integrated extrusion-based 3D food printer and a graphical user interface (GUI) with a custom G-code generator for assigning taste categories and intensities across layers and spatial locations within a single print.
We evaluate TastePrint through technical studies of spray resolution and deposited mass, a minimal sensory discrimination study on taste-pattern perception, and an exploratory formative user-feedback study on GUI use. The primary objective of this paper is to establish the technical feasibility of layer-wise spatial taste design within the tested workflow.}


\section{Materials and Methods}
\label{materials_and_methods}
\subsection{\revdel{Experimental Equipment: TastePrint System}\revadd{TastePrint System Overview}}
\label{system_overview}
\revadd{The TastePrint system comprised a spray-integrated extrusion-based 3D food printer and a graphical user interface (GUI) for layer-wise taste design and G-code generation.
The following subsections describe the overall workflow, hardware configuration, interface, and synchronized fabrication process.}

\subsubsection{Workflow of TastePrint}
\label{printing_process}

The TastePrint fabrication process proceeds as follows (Figure~\ref{fig:system_overview}):
\begin{enumerate}
    \item Prepare a 3D model of the desired food geometry in a standard format (e.g., STL, OBJ).
    \item Import the model into the GUI, slice it into layers, and assign taste distributions for each layer.
    \item The system automatically calculates spray parameters--including nozzle position, spray duration, and intensity--based on user inputs.
    \item Export a modified G-code file that integrates both extrusion and spray commands.
    \item During printing, the syringe extrudes the base food ink while the system sprays the designated liquid seasonings (e.g., salt solutions) at specified positions.
    \item After printing, the food is removed from the platform for immediate consumption or optional post-processing such as baking or frying.
\end{enumerate}

\begin{figure*}[t]
    \centering
    \includegraphics[width=\textwidth]{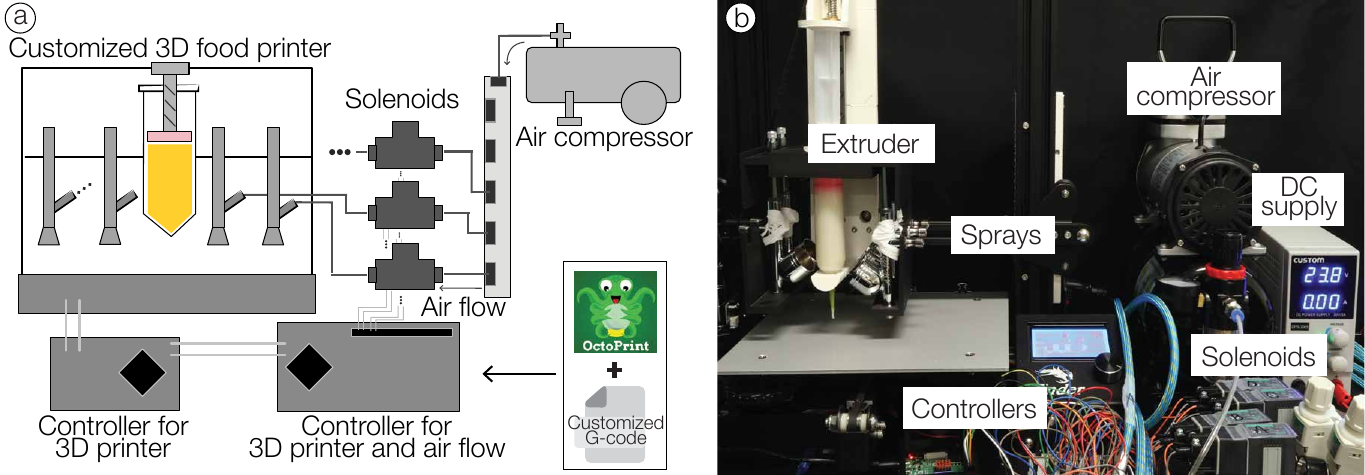}
    \caption{System overview of the spray-integrated 3D food printer: (a) Schematic diagram of the customized 3D food printer integrating air-based spraying and extrusion. The system consists of solenoid valves driven by a DC power supply and controlled via OctoPrint through customized G-code commands. The air compressor provides pressurized airflow to the solenoids, which regulate spray activation in synchronization with the 3D printer controller. (b) Photograph of the actual setup showing the key components, including the extruder, spray units, solenoids, air compressor, DC supply, and microcontroller modules that coordinate printing and spraying.
    }
    \label{fig:printer_setup}
\end{figure*}

\subsubsection{Customization of 3D Food Printer}
\label{printer}
The customized printer (Figure~\ref{fig:printer_setup}) is based on a commercially available FDM 3D printer (Ender 3, Creality, China) modified for food printing.
The thermoplastic extruder was replaced with a syringe-based extrusion unit (30 mL capacity, 1.6 mm nozzle) for dispensing food inks, and an airbrush holder was installed adjacent to the nozzle for seasoning application.
The firmware was replaced with a custom Marlin-based version supporting syringe control and extended G-code commands for airbrush activation.
These commands communicate with a Raspberry Pi 4\footnote{\url{https://www.raspberrypi.com/products/raspberry-pi-4-model-b/}} 
running OctoPrint\footnote{\url{https://octoprint.org/}} for synchronized control of extrusion and spraying.
Food-contact components and sanitization procedures are described in Section~\ref{sanitization}.

\begin{figure}[t]
    \centering
    \includegraphics[width=0.5\columnwidth]{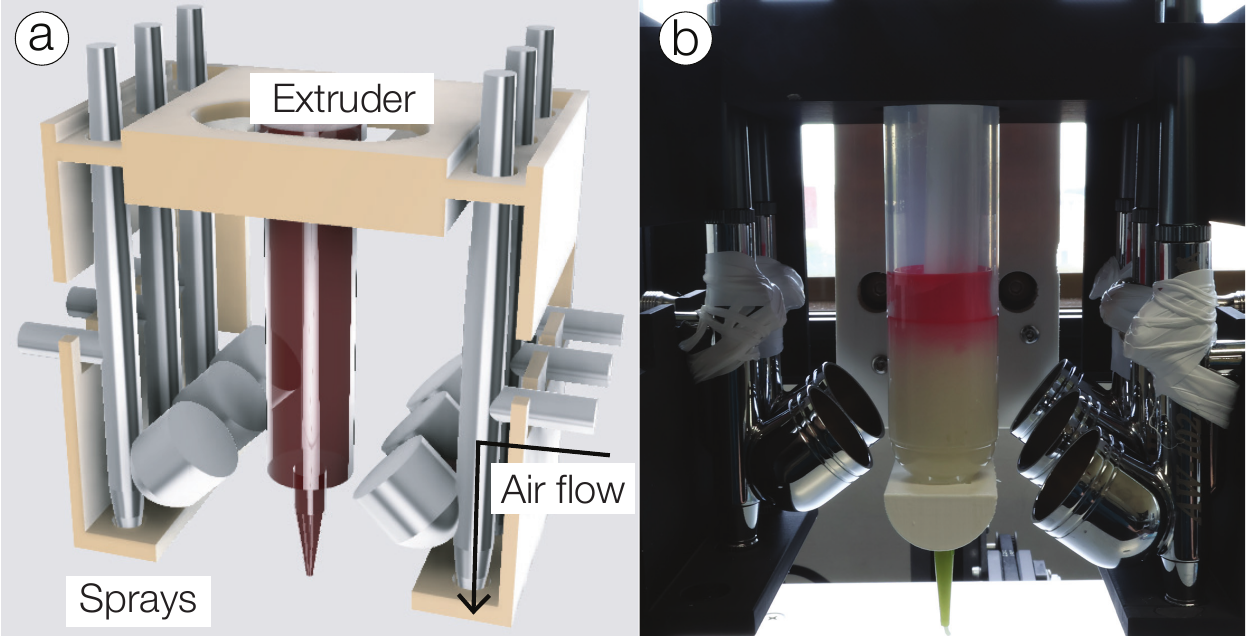}
    \caption{Printer head design: (a) Schematic design showing one material extruder and six spray holders for seasoning. (b) Fabricated printer head integrated into the 3D food printer.
    }
    \label{fig:printer_head_design}
\end{figure}

\subsubsection{Airbrushing Mechanism}
\label{airbrush}
The airbrushing mechanism, mounted adjacent to the print head (Figure~\ref{fig:printer_head_design}), enables spatially controlled deposition of liquid seasonings during printing.
The prototype employs six independently controlled airbrushes (AW-102, Airbrush Works) arranged radially around the extruder, each with a nominal 0.2 mm nozzle.
Each airbrush connects to a miniature solenoid valve (FFBM-2106A, CKD) supplied by an air compressor (Ausuc) and pressure regulator (RB500, CKD).
The solenoids are powered by a 24 V DC supply (DPS-3005, CUSTOM) and switched via GPIO pins on the Raspberry Pi 4.

Spray duration and timing are fully programmable through the control software, enabling fine adjustments to taste intensity. 
Custom G-code commands synchronize valve activation with the extrusion path and layer transitions, ensuring precise alignment between geometry formation and seasoning deposition.

\subsubsection{Graphical User Interface}
\label{gui}
\begin{figure}[t]
    \centering
    \includegraphics[width=0.5\columnwidth]{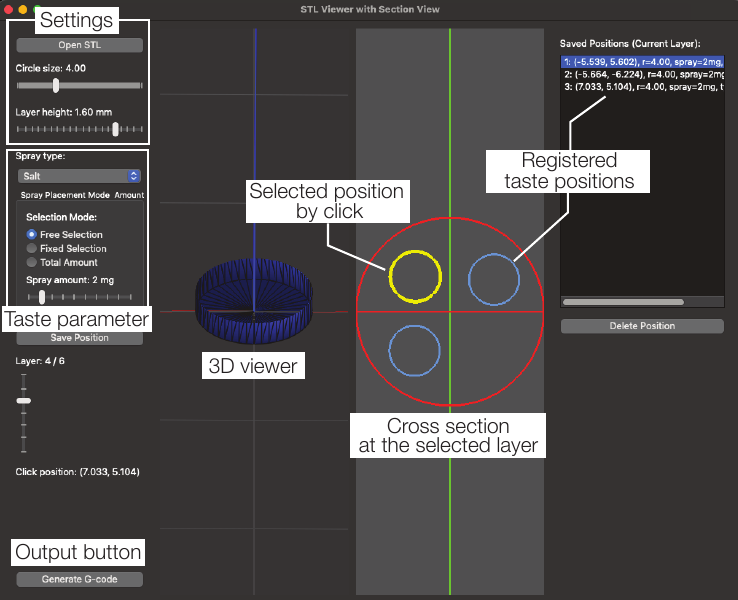}
    \caption{Interface of the spray-position editing tool: The interface enables users to specify where to spray flavors within each layer of a 3D food model. The left panel provides settings for model loading, layer height, spray parameters, and output generation. The central viewer displays the 3D model and its cross-section at the selected layer, allowing users to click and register desired taste positions. The registered positions are shown as colored markers on the cross-section and listed on the right side of the interface.
    }
    \label{fig:gui}
\end{figure}

\revadd{The TastePrint GUI, implemented in Python 3.11 using PyQt5 (Qt 5.15), provides an interactive environment for designing layer-wise and spatial taste distributions (Figure~\ref{fig:gui}).
Users import a 3D food model (STL or OBJ), specify extrusion parameters such as layer height and nozzle diameter, and optionally define infill patterns.
The system slices the model into layers based on these settings, after which users assign taste distributions by selecting seasoning channels, spray positions, and spray intensities for each layer, enabling different seasoning types and concentration levels to be combined within a single fabricated item.}
The GUI offers three modes of taste design:
\begin{itemize}
    \item \textbf{Free selection mode}: Users manually select spray positions and intensities for each layer.
    \item \textbf{Patterned selection mode}: Predefined spatial templates (e.g., dense packing) assist rapid pattern generation.
    \item \textbf{Total amount mode}: Users specify the overall amount of each taste for the entire model; the system automatically allocates spray amounts across layers according to geometry and user-defined constraints.
\end{itemize}
These modes accommodate different user needs: professional chefs may prefer direct, detailed control, whereas casual users can rely on automatic allocation.
Finally, the GUI exports a custom G-code file that integrates extrusion and spray commands for fabrication. This interface bridges digital food modeling and hardware control, allowing users to receive a reproducible fabrication file that encodes their designed taste distribution.

\subsubsection{3D food printing with seasoning spray}
\label{printing_process_details}
\begin{figure}[t]
    \centering
    \includegraphics[width=\columnwidth]{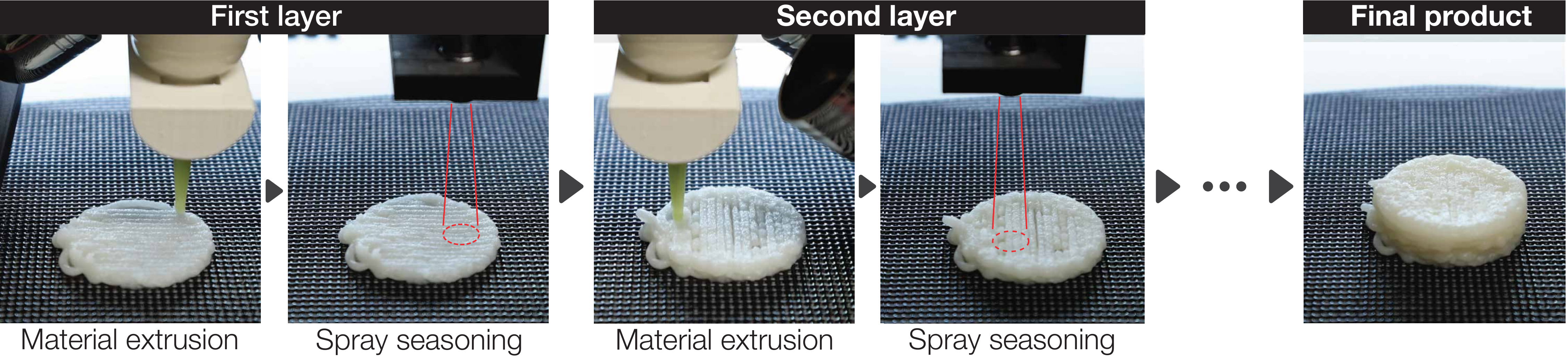}
    \caption{
        Sequential fabrication process of the TastePrint system.
        }
    \label{fig:print_result}
\end{figure}

Printing proceeds in a layer-by-layer workflow following the modified G-code (Figure~\ref{fig:print_result}).
For each layer, the printer first extrudes the base food ink according to the defined geometry.
\revdel{Upon completion, the print head moves to the designated spray positions, where the corresponding airbrushes are activated for specified durations to deposit the selected liquid seasonings.}
\revadd{Upon completion, the print head moves to one or more designated spray positions, where the selected seasoning channels are activated for programmed durations to deposit liquid seasonings.
Each layer can therefore include multiple spray events at different locations with different seasoning assignments.}

Extrusion and spraying are synchronized through custom G-code commands, ensuring that the taste application aligns precisely with each printed layer.
This process repeats until fabrication is complete.
The resulting food items exhibit spatially controlled taste distributions and can be consumed directly or subjected to optional post-processing such as baking or frying.

\subsection{Experimental Materials}
\label{experimental_materials}
\paragraph{Base Food Material}
Mashed potato paste was selected as the base material because its rheological properties are well-suited for extrusion-based 3D food printing.
The paste was prepared by mixing commercially available instant mashed potato flakes (Nichiga, Japan) with hot water at a weight ratio of 4:1 (water : flakes) and stirring until a homogeneous consistency was achieved.
The resulting mixture exhibited stable extrusion behavior and smooth layer formation.
The prepared paste was loaded into a 30 mL disposable syringe (Nordson) for printing. Mashed potato was used in this study for reproducibility.

\paragraph{Seasoning Liquid}
Up to six airbrushes can be mounted on the TastePrint system; five were used in this study to represent the basic tastes—sweet, salty, sour, bitter, and umami—while one nozzle was reserved for future extensions. The corresponding seasoning agents were sucrose (sweet), sodium chloride (salty), citric acid (sour), caffeine (bitter), and monosodium glutamate (umami).
Concentrations were selected through internal pilot trials to achieve perceptible yet balanced intensities when sprayed onto the base material, and can be adjusted according to the intended application.

To improve atomization and prevent nozzle dripping, xanthan gum was added to each solution at 0.5 wt\%.
This viscosity adjustment produced stable, uniform spray patterns across all taste channels, as judged from the absence of dripping and from repeatable spray footprints under the tested conditions.
Food-grade coloring agents were incorporated into each solution solely for visualization and quantitative analysis of spray distributions.

\subsection{Sanitization Procedures}
\label{sanitization}
A standardized sanitization protocol was followed to ensure hygienic fabrication and sensory testing.
All reusable food-contact surfaces were washed with detergent, rinsed, and wiped with 70\% ethanol before each printing session.
Components not rated as food-safe were kept isolated from edible contact.
Airbrush tanks and seasoning lines were flushed with ethanol between sessions to remove residue and minimize microbial growth.

Syringes and extrusion nozzles were single-use disposables and discarded after each print. To prevent cross-contamination, each seasoning was assigned a dedicated reservoir and tubing line, which were flushed sequentially with sterile water and ethanol, and then air-dried before reuse on the same day. Operators wore gloves and masks during all preparation and printing steps, replacing gloves between different formulations.

Prepared food materials were stored below 10 °C or used within 2 h at room temperature, and any remaining materials were discarded. This procedure ensured hygienic operation and repeatability of both printing and tasting experiments.

This protocol reduces microbial risk and cross-contamination while maintaining the repeatability of the printing and tasting procedures.

\subsection{Evaluation}
\label{evaluation}
The TastePrint system was evaluated through two technical experiments and two human-subject studies: (1) spray resolution, (2) spray amount per shot, (3) a minimal sensory discrimination study on taste localization, and (4) an exploratory formative user-feedback study after a design-and-print task.
The technical experiments quantified the precision and controllability of the airbrushing mechanism, whereas the human-subject studies provided preliminary observations on taste-pattern perception and on how non-expert users interpreted the layer-wise taste-design workflow.

\subsubsection{Spray resolution}
\label{printing_accuracy}
We conducted a spray test to quantify the spatial precision of liquid-seasoning deposition as a function of nozzle-to-surface distance and spray duration.

\paragraph{Setup}
Dyed seasoning liquid was first sprayed onto filter paper affixed to the printer platform to obtain a repeatable imaging-based calibration of the spray footprint.
ArUco markers (20 mm) were placed around the target to enable pixel-to-millimeter calibration and camera pose estimation.

\paragraph{Test pattern and factors}
Distances of 20, 30, and 40 mm and spray durations of 20, 40, and 60 ms were tested at a line pressure of 0.10 MPa, with three replicates per condition (n = 27).
Images were captured under controlled lighting using a top-mounted camera.

\paragraph{Imaging, image processing, and metrics}
Images were rectified using the computed homography matrix, and the sprayed region was segmented by Otsu thresholding on the red colour channel.
The largest connected component within a 24 × 24 mm ROI was extracted to compute the equivalent circular diameter of the spray spot.

\paragraph{Model fitting}
A linear regression model was fitted:
\begin{align}
    \text{Spray Size (mm)} =~ &\beta_0 + \beta_1 \times \sqrt{\text{Distance (mm)}}~+ \nonumber\\ 
    &\beta_2 \times \sqrt{\text{Duration (ms)}}
\end{align}
where $\beta_0$ is the intercept, $\beta_1$ and $\beta_2$ are coefficients for squared distance and duration, respectively.
The square-root terms accounted for the non-linear spread behaviour observed in preliminary trials.

\paragraph{Edible-substrate transfer check}
\revadd{
To examine whether the calibration trend obtained on filter paper transfers to an edible substrate, we repeated representative spray conditions on printed mashed-potato samples.
Circular samples (30 mm diameter, 4.8 mm height) were fabricated using the same mashed-potato formulation and printing settings as in the main experiments.
Dyed seasoning liquid was sprayed once at the centre of each sample at nozzle-to-surface distances of 20 and 30 mm and spray durations of 20 and 60 ms under a line pressure of 0.10 MPa, with three replicates per condition.
Top-view images were captured immediately after spraying and again after 5 min under fixed lighting with a calibration marker.
Using the same rectification and segmentation pipeline as in the filter-paper experiment, we computed the equivalent circular diameter of the stained region and quantified short-term lateral spread as the increase in diameter over 5 min.}

\subsubsection{Spray amount per shot}
\label{spray_amount_per_shot}
We also conducted a second spray test to measure the quantity of seasoning deposited per spray activation and determine its relationship with spray duration.
\paragraph{Experimental setup and protocol}
A digital micro-scale (Fincy Palmoo; 0.001 g resolution) was placed on the printer platform. Liquid seasoning was sprayed directly onto the scale at a fixed distance of 20 mm and pressure of 0.10 MPa. Durations of 10, 20, 40, 60, and 80 ms were tested, with three replicates each (n = 15).

\paragraph{Analysis}
Mass change after each activation was recorded after the reading stabilized.
The relationship between spray duration and deposited mass was modelled as:
\begin{equation}
    \text{Spray Amount (mg)} = \alpha_0 + \alpha_1 \times \text{Duration (ms)}
\end{equation}
where $\alpha_0$ is the intercept and $\alpha_1$ is the coefficient for duration.
Regression parameters and coefficients of determination were used to assess linearity.

\subsubsection{Sensory discrimination study}
\label{taste_localization_study}
\revadd{We conducted a small sensory discrimination study to examine whether a spatially concentrated seasoning pattern could be perceptually distinguished from a more distributed one under controlled tasting conditions.
The study provided an initial perceptual assessment of taste localization using printed samples.}

\begin{figure}[t]
    \centering
    \includegraphics[width=0.5\columnwidth]{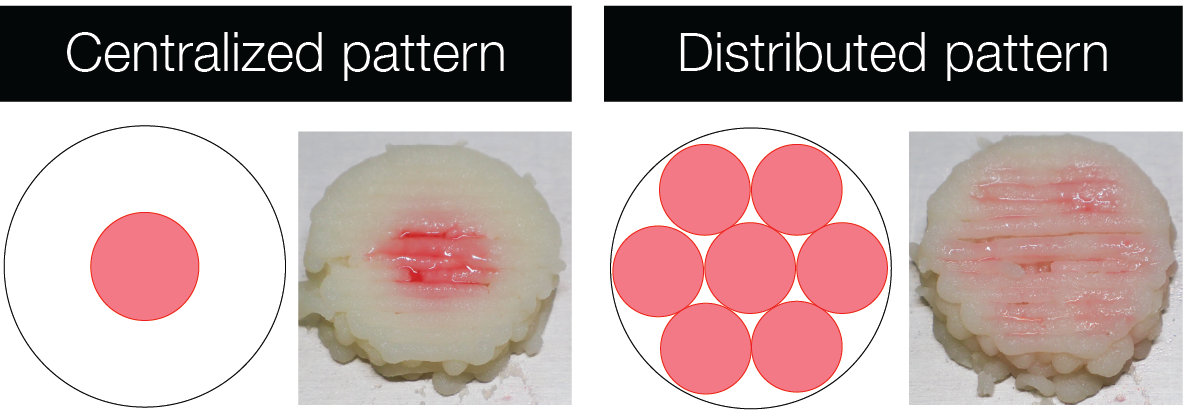}
\caption{Schematic and dyed-sample visualization of the seasoning patterns used in the sensory discrimination study. The centralized pattern is shown on the left and the distributed pattern on the right; for each condition, the schematic is followed by a dyed sample used only to visualize the intended spatial distribution. Clear seasoning solutions were used in the actual sensory study so that the samples were visually indistinguishable.}
    \label{fig:user_study_stimulus}
\end{figure}

\paragraph{Participants and stimuli}
\revadd{
Ten participants (two females, eight males) completed the study.
Two mashed-potato samples with identical geometry and nominal total seasoning amount were prepared for each trial: a \emph{centralized} condition, in which the salty seasoning was concentrated at the center region, and a \emph{distributed} condition, in which the salty seasoning was spread more broadly across the sample.
All samples were fabricated using the sam mashed-potato base and the same seasoning channel, and were visually indistinguishable due to the use of clear seasoning solutions.}
\revadd{Figure~\ref{fig:user_study_stimulus} illustrates the centralized and distributed seasoning patterns used in the sensory discrimination study. For each condition, a schematic and a dyed sample are shown to visualize the intended spatial distribution.}

\paragraph{Procedure}
\revadd{
The experiment followed a within-subject two-alternative forced-choice design.
Each participant completed four trials.
In each trial, participants were presented with one centralized and one distributed sample in randomized left-right order and answered which sample felt more localized in taste.
To prevent fatigue and maintain sensitivity, participants were instructed to cleanse their palate with water between each consumption.
This study protocol was approved by the Institutional Review Board (No. R6-E-51) of the institution, and informed consent was obtained from all participants.}

\paragraph{Analysis}
\revadd{
We calculated the proportion of trials in which participants selected the centralized sample and compared it against the 50\% chance level expected in a two-alternative forced-choice task.}

\subsubsection{\revdel{Usability of TastePrint System}\revadd{Exploratory formative user-feedback study}}
\label{user_feedback_study}
We conducted \revdel{a user study}\revadd{an exploratory formative user-feedback study} to assess how first-time users interpreted and used the TastePrint interface in a design-and-print task.

\paragraph{Participants and task}
Three home cooks (one female, two males) with no prior experience in 3D food printing used the GUI to design and fabricate a food item with spatially varied tastes.

\paragraph{Procedure}
Participants completed a single design-and-print session (~30 min) and were then interviewed in a semi-structured format.
Questions focused on (1) ease of use, (2) clarity of layer-wise taste design, and (3) perceived applications.

\section{Results and Discussion}
\label{results_and_discussion}
This section reports the performance of the TastePrint system in terms of spray stability, deposition accuracy, preliminary perceptual discrimination, and exploratory user feedback. Results are presented alongside interpretive discussion to contextualize what the current data support for layer-wise taste control in 3D-printed foods.  

\subsection{Seasoning liquids}
\label{seasoning_liquids_results}
\revadd{
All seasoning aqueous solutions with 0.5 wt\% xanthan gum were sprayed successfully without clogging or instability across the 30 trials for each solution.
The addition of 0.5 wt\% xanthan gum helped stabilize the sprayed liquids and produced uniform atomization.
In contrast, solutions without xanthan gum exhibited nozzle dripping during idle periods and irregular spray patterns.
This dripping likely occurred because the airbrush nozzles remained mechanically open and spray timing was controlled upstream by solenoid valves in the present setup; under this configuration, lower-viscosity liquids were more prone to dripping from the nozzle tip during idle periods.
Accordingly, increasing viscosity with xanthan gum was an effective strategy for mitigating dripping and achieving stable spraying.
This result indicates that, in the present upstream-valve architecture, seasoning formulation is not merely a material parameter but part of the control design space, because spray stability depends jointly on fluid viscosity and nozzle-state management.}

\subsection{Effect of layer height for taste spraying}
\label{layer_height_results}
The mashed-potato paste extruded smoothly at a nominal layer height of 1.6 mm.
Within the current workflow, spray duration can be scaled proportionally with layer height to maintain nominal seasoning dosage per unit height.
The GUI can automate this adjustment across different vertical resolutions.

\subsection{Spray resolution}
\label{printing_accuracy_results}
\begin{figure}[t]
    \centering
    \includegraphics[width=0.7\columnwidth]{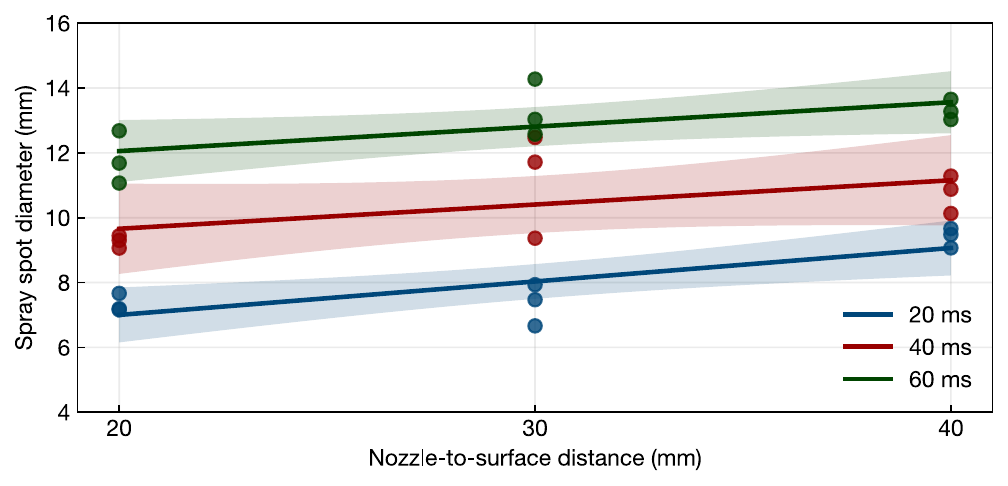}
    \caption{Results of the spray-resolution experiment. Spray spot diameter is plotted against nozzle-to-surface distance, with separate series for spray durations of 20, 40, and 60 ms. Points indicate individual measured spray diameters, solid lines indicate the fitted regression model for each duration, and shaded bands indicate the 95\% confidence intervals of the fitted mean.}
    \label{fig:spray_resolution}
\end{figure}

Figure~\ref{fig:spray_resolution} shows the measured relationship between spray spot diameter, nozzle-to-surface distance, and spray duration. Both factors increased spray diameter, consistent with fluid-dynamic expectations. The fitted regression model achieved a coefficient of determination of R$^2$ = 0.86, indicating good predictive accuracy.
The mean standard deviation across replicates was 0.79 mm, demonstrating reliable repeatability.
The coefficients were $\beta_0 = -3.525$, $\beta_1 = 1.450$, and $\beta_2 = 0.918$.
These results verify that users can achieve desired spatial resolutions by tuning distance and spray duration—critical parameters for designing fine-grained taste layouts.

\begin{figure}[t]
    \centering
    \includegraphics[width=0.7\columnwidth]{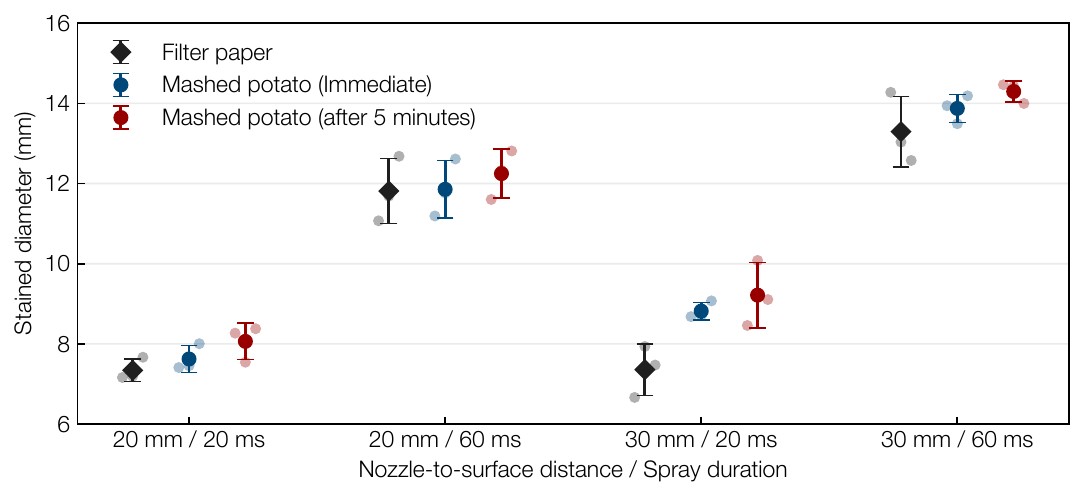}
    \caption{Comparison of stained diameters measured on filter paper and on printed mashed-potato samples under representative spray conditions. For each nozzle-to-surface distance and spray-duration combination, black markers indicate filter-paper measurements, blue markers indicate mashed-potato measurements taken immediately after spraying, and red markers indicate mashed-potato measurements taken 5 min after spraying. Large markers and error bars indicate mean $\pm$ standard deviation, and faint points indicate individual replicates.}
    \label{fig:spray_resolution_mashpotato}
\end{figure}

\revadd{The edible-substrate transfer check showed that the stained diameters measured on mashed-potato samples immediately after spraying were broadly comparable to those measured on filter paper under the corresponding conditions (Figure~\ref{fig:spray_resolution_mashpotato}).
A noticeable deviation was observed in the 30 mm distance and 20 ms duration condition, indicating that filter paper is useful for first-pass calibration of spray conditions, while final tuning for food fabrication benefits from measurements on the target edible substrate.}
\revadd{Over 5 min after spraying, the stained diameter on mashed-potato samples increased by approximately 4\% on average, indicating modest lateral spread on the edible substrate.
Because this increase was small relative to the initial diameter, spatial localization was largely preserved over the short time interval tested.}

\subsection{Spray amount per shot}
\label{spray_amount_per_shot_results}
\begin{figure}[t]
    \centering
    \includegraphics[width=0.7\columnwidth]{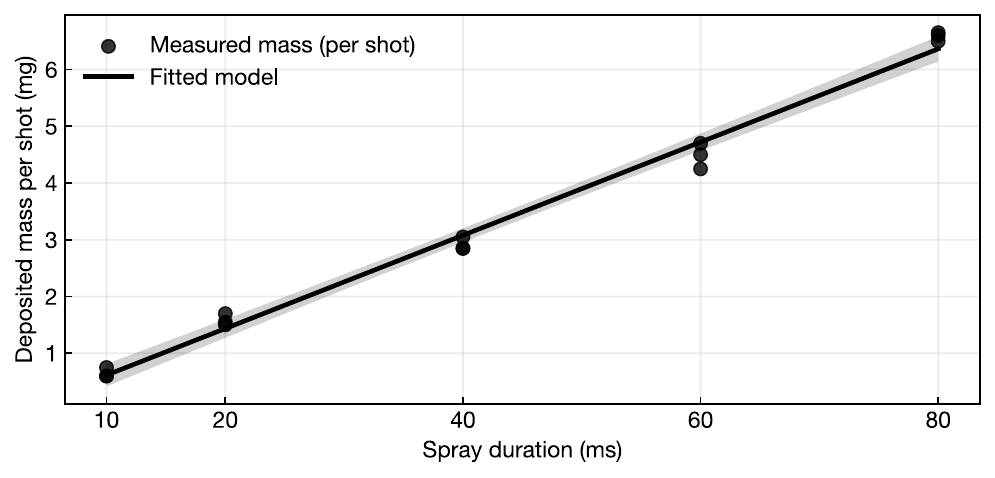}
    \caption{Results of the spray-amount-per-shot experiment. Points indicate individual measured mass values, the solid line shows the fitted linear regression model, and the shaded band shows the 95\% confidence interval.}
    \label{fig:spray_amount}
\end{figure}

The amount of deposited liquid increased linearly with spray duration, R$^2$ = 0.99, confirming highly predictable deposition. The average standard deviation was 0.2 mg, indicating precise control of dispensed volume, as shown in Figure~\ref{fig:spray_amount}.
\revdel{The coefficients were $\alpha_0 = 0.082$ and $\alpha_1 = -0.206$.}
\revadd{The coefficients were $\alpha_0 = -0.206$ and $\alpha_1 = 0.082$.}
This linear controllability allows direct mapping between GUI-specified spray duration and delivered seasoning quantity, enabling accurate quantitative taste modulation.
\revadd{This relationship allows the deposited seasoning amount to be estimated directly from spray duration, supporting quantitative control of seasoning application at the hardware level.}

\subsection{Controllability of seasoning}
\label{controllability_of_seasoning_results}
Combining the resolution and deposition results shows that TastePrint can decouple spatial placement from nominal dosage assignment at the hardware-control level.
Because deposited mass can be estimated from spray duration and solution concentration, the system provides a practical basis for specifying seasoning patterns in terms of both geometry and quantity during fabrication.
\revadd{Together, these results provide a technical basis for subsequent food-substrate and perceptual validation of spatial taste-pattern design.}

\subsection{Sensory discrimination study}
\revadd{
Across all answered trials, participants identified the centralized sample as more localized in taste in 27 of 40 trials (67.5\%).
At the participant level, responses were mixed: some participants consistently selected the centralized sample, whereas others did not show a clear preference across the four trials.
These results suggest a tendency for the centralized seasoning pattern to be perceived as more spatially localized under the tested conditions, although the effect was not uniform across participants.
They provide initial evidence that spatial taste arrangement can function as a meaningful design variable.}

\subsection{\revdel{Usability of TastePrint System}\revadd{Exploratory Usability Feedback}}
\label{sensory_evaluation_results}
All three participants successfully completed the assigned design-and-print session and produced foods with spatially varied taste layouts.
Average task completion time was approximately 15 minutes. Interviews suggested that the GUI was understandable even for novices and that the layer-by-layer cross-sectional view helped participants reason about internal taste placement.
Participants highlighted the system's potential for personalized meals, aesthetic food design, and creative experimentation with taste combinations.

Constructive feedback included requests for additional taste channels, more realistic food rendering, and a preview of final taste distribution before printing.
These suggestions identify priorities for future software iterations.
Overall, this exploratory feedback suggests that TastePrint can support early-stage end-user interaction design.
\revadd{The study suggests that participants were able to complete a guided taste-design task and articulate concrete improvement requests.}

\subsection{Limitations}
\label{limitations_and_future_perspectives}
While TastePrint demonstrates the feasibility of layer-wise taste modulation in 3D food printing, several limitations remain.

\paragraph{Process and Channel Constraints}
In the current implementation, seasoning is applied sequentially after each printed layer, increasing total fabrication time relative to extrusion-only workflows.
In addition, the prototype supports six airbrush channels, each corresponding to one seasoning liquid, which restricts the achievable complexity of taste combinations.

\paragraph{Material, Rheology, and Spray Stability}
\revadd{The current validation focused on a single mashed-potato formulation, which has favorable absorption and surface properties for spray deposition. Less porous materials, such as meat-based inks, may exhibit poorer adhesion or greater lateral diffusion of sprayed seasonings.Likewise, although xanthan-gum supplementation improved spray stability in the tested formulations, the present study did not quantify viscosity or broader rheological operating bounds across formulations.
In our current setup, increasing the xanthan gum concentration to 0.7 wt\% led to clogging, suggesting that the practical operating window is bounded not only by formulation rheology but also by the spray hardware configuration, including nozzle size and actuation conditions.
In addition, the current nozzle configuration required viscosity adjustment to avoid dripping during idle periods, and broader formulation compatibility was not systematically evaluated.}

\paragraph{Structural Stability and Temporal Spread}
\revadd{The spray-resolution model was calibrated on filter paper, and although the resulting footprint estimates were broadly consistent with those observed on mashed-potato samples immediately after spraying, the model should still be interpreted as an empirical operating model for the present setup rather than as a material-independent footprint law.
The mashed-potato transfer check showed a modest increase in stained diameter over 5 min, indicating short-term lateral spread within the food matrix. Accordingly, the long-term stability of spatial taste patterns, especially during extended storage, should not be assumed from the present results alone.
Smaller layer heights can, in principle, provide finer control of taste distribution along the vertical axis because seasoning is assigned layer by layer. Under the current setup, fabricated samples retained their overall shape during repeated spraying in our fabrication trials; for example, cylindrical samples (5 mm height, 25 mm diameter) retained their shape after 50 spray events of 20 ms each, suggesting that gross collapse was not observed under the tested conditions. However, very thin layers may still be more susceptible to disturbance from spray airflow and deposited liquid, and the present study did not directly quantify deformation or failure thresholds. Layer-height selection should therefore balance vertical taste resolution against printability and structural stability within the current workflow.}

\paragraph{Perceptual and Scope Constraints}
\revadd{
The current sensory evidence is limited to a minimal discrimination study under the reported conditions, and broader validation is still needed to determine how robustly spatial seasoning patterns remain perceivable across different foods, concentration settings, bite sequences, and user populations.}
Likewise, deposited mass should not be equated directly with perceived intensity, because taste perception also depends on dissolution, local concentration, oral mixing, and individual sensitivity. Finally, TastePrint presently manipulates gustatory stimuli (sweet, salty, sour, bitter, umami) but does not address aroma or retronasal flavor, which are central to holistic food perception.

\section{Conclusion}
\label{conclusion}
This study introduced TastePrint, a 3D food printing system that enables layer-wise spatial control of taste through programmable airbrushing of liquid seasonings.
Integrating a custom graphical interface with synchronized hardware control, the system enables complex internal taste architectures using a single base material.

\revadd{Technical evaluations showed high spatial precision and predictable controllability of seasoning deposition, while the edible-substrate transfer check showed broad consistency between filter-paper calibration and measurements on printed mashed-potato samples. A minimal sensory discrimination study provided initial evidence that spatial taste arrangement can remain perceptually meaningful after fabrication, and the exploratory formative user-feedback study suggested that participants were able to complete a guided taste-design task and articulate concrete improvement requests. Together, these results support the technical feasibility of separating geometric fabrication from taste modulation as a framework for digital food design.

TastePrint therefore establishes a technical foundation for computational taste-geometry co-design within the tested mashed-potato formulation and current spray-based workflow.
Further validation across food materials, broader sensory conditions, and more complex three-dimensional taste layouts will be important for extending the framework beyond the present prototype.}

\section*{Acknowledgments}
This work was supported by JST ACT-X Grant Number JPMJAX24CQ, and JSPS KAKENHI Grant Number 23K11198, Japan.

\section*{\revadd{CRediT authorship contribution statement}}
\revadd{Yamato Miyatake: Conceptualization, Methodology, Software, Hardware integration, Investigation, Formal analysis, Visualization, Writing - original draft. Parinya Punpongsanon: Supervision, Funding acquisition, Project administration, Writing - review \& editing.}

\section*{Declaration of generative AI and AI-assisted technologies in the manuscript preparation process}
During the preparation of this work, the author(s) used ChatGPT in order to improve the clarity and coherence of the text. After using this tool/service, the author(s) reviewed and edited the content as needed and take(s) full responsibility for the content of the published article.
\bibliographystyle{elsarticle-harv}
\bibliography{cas-refs}
\end{document}